\documentclass[11pt,prd,preprint,superscriptaddress]{revtex4}
 \pdfoutput=1
\usepackage{revsymb}
\usepackage{amssymb}
\usepackage{amsmath}
\usepackage[pdftex]{graphicx}
\usepackage{bm}
\usepackage[utf8]{inputenc}
\usepackage{float}
\usepackage{amsfonts}
\usepackage{placeins}
\newcommand{\be}{\begin{equation}}
\newcommand{\ee}{\end{equation}}
\newcommand{\ben}{\begin{eqnarray}}
\newcommand{\een}{\end{eqnarray}}

\begin{document}
\title{Averaged Lema\^{\i}tre-Tolman-Bondi dynamics}
%\date{\today}

\author{Eddy~G.~Chirinos Isidro\footnote{E-mail: eddychirinos.isidro@cosmo-ufes.org}}
\affiliation{Universidade Federal do Esp\'{\i}rito Santo,
Departamento
de F\'{\i}sica\\
Av. Fernando Ferrari, 514, Campus de Goiabeiras, CEP 29075-910,
Vit\'oria, Esp\'{\i}rito Santo, Brazil}
\author{Rodrigo M. Barbosa
%\footnote{E-mail: rodrigo_martins@email.com}
}
\affiliation{Universidade Federal do Esp\'{\i}rito Santo,
Departamento
de F\'{\i}sica\\
Av. Fernando Ferrari, 514, Campus de Goiabeiras, CEP 29075-910,
Vit\'oria, Esp\'{\i}rito Santo, Brazil}

\author{Oliver~F.~Piattella\footnote{E-mail: oliver.piattella@pq.cnpq.br}}
\affiliation{Universidade Federal do Esp\'{\i}rito Santo,
Departamento
de F\'{\i}sica\\
Av. Fernando Ferrari, 514, Campus de Goiabeiras, CEP 29075-910,
Vit\'oria, Esp\'{\i}rito Santo, Brazil}

\author{Winfried Zimdahl\footnote{E-mail: winfried.zimdahl@pq.cnpq.br}}
\affiliation{Universidade Federal do Esp\'{\i}rito Santo,
Departamento
de F\'{\i}sica\\
Av. Fernando Ferrari, 514, Campus de Goiabeiras, CEP 29075-910,
Vit\'oria, Esp\'{\i}rito Santo, Brazil}

\date{\today}

%\date{\today}
\begin{abstract}
We consider cosmological backreaction effects in Buchert's averaging formalism on the basis of an explicit solution of the Lema\^{\i}tre-Tolman-Bondi (LTB) dynamics which is linear in the LTB curvature parameter and has an inhomogeneous bang time. The volume Hubble rate is found in terms of the volume scale factor which represents a derivation of the simplest phenomenological solution of Buchert's equations in which the fractional densities corresponding to  average curvature and kinematic backreaction are explicitly determined by the parameters of the underlying LTB solution at the boundary of the averaging volume. This configuration represents an exactly solvable toy model but it does not adequately describe our ``real" Universe.
%averaged curvature and the kinematic backreaction
\end{abstract}
%\pacs{98.80.-k, 95.35.+d, 95.36.+x}

\maketitle

\section{Introduction}

The cosmological standard model is based on the cosmological principle according to which our Universe is spatially homogeneous and isotropic at large scales. The value of the corresponding homogeneity scale is a matter of debate but it is assumed to be about one order of magnitude smaller than the size of the observable Universe.
On smaller scales, scales of galaxy clusters and below, the cosmos is clearly inhomogeneous.
Mathematically, the cosmological principle implies the existence of time orthogonal subspaces of constant curvature, i.e., it is characterized by true symmetries of the spacetime.
A possibly more realistic view is to regard the
Universe as only statistically homogeneous and isotropic as a result of a suitable averaging procedure over inhomogeneities and without assuming a fictitious highly symmetric background.
How to perform averages in General Relativity (GR) and particularly in cosmology is not yet a really established issue. Nevertheless, there are well motivated approaches which are believed to capture essential features of  the problem \cite{EllisStoeger:fitting 1987,EllisCQG,WiltshireCQG,BuchertCQG,RasaCQG,KolbCQG}.
A very general manner to deal with the averaging problem relies on the exact covariant Macroscopic Gravity
formalism by Zalaletdinov \cite{zala}. Macroscopic gravity has been applied to spherically symmetric cosmology by Coley et al. in \cite{coley,coley2,coley3}, observational aspects of the resulting dynamics were discussed in \cite{coleyobs}. A result of this averaging is the appearance of an additional spatial curvature term in the dynamical equations.
Restricting ourselves to averages over scalar quantities, we shall make use here of Buchert's approach \cite{Buchert:1999er} which will allow for an explicit calculation of the averages of interest in the context of this paper.

The structures in the Universe are characterized by different length scales: the scale of the solar system, the scale of galaxies, the scale of galaxy clusters and the homogeneity scale.
It is believed that the dynamics on a larger scale may, in principle, be obtained by an averaging procedure over the dynamics on the underlying smaller scale in case this dynamics is known.
What one would like to have is an explicit connection between two different levels, characterized by length scales $l_{1}$ and $l_{2}$ with  $l_{2}\gg l_{1}$, where, in particular, $l_{2}$ may be of the order of the homogeneity scale.
Assuming the cosmological evolution to be governed by GR, the question arises, on which scale Einstein's equations will be valid. The safest starting point may be the solar-system scale since the validity of GR on this scale is established with high precision.
Generally, if we consider GR to be valid on a scale $l_{1}$ then, because of its nonlinearity, it is certainly no longer valid in its conventional form after an averaging procedure which is expected to account for the physics on a scale $l_{2}\gg l_{1}$.
In particular, this is true if we consider the transition to a spatially homogeneous description at a scale $l \gtrsim l_{2}$ from an underlying inhomogeneous one for $l \approx l_{1}$.
The Buchert equations are an approach to perform this step \cite{Buchert:1999er}. Starting from an irrotational pressureless matter distribution, the spatial average of the inhomogeneous dynamics over a certain rest mass preserving domain $D$ results in a set of equations for the domain dependent volume scale factor $a_{D}(t)$ which depends on time and on the parameters of the domain $D$.
This quantity  $a_{D}(t)$ is defined through the time dependent domain volume $V_{D}(t)$ by $a_{D} \propto V_{D}^{1/3}$.
The equations for $a_{D}(t)$ have the structure of the Friedmann and acceleration equations of standard cosmology but they do not rely on a homogeneous and isotropic background and  $a_{D}(t)$ is not the scale factor of a  Robertson-Walker (RW) metric since, through the averaging operation, the scale factor must be scale-dependent.
However, compared with the equations for a Friedmann-Lema\^{\i}tre-Robertson-Walker (FLRW) universe there appear additional terms as a consequence of the averaging process: a kinematical backreaction and an averaged curvature term. These terms are related by a consistency condition. In their absence one recovers the dynamics of an Einstein-de Sitter (EdS) universe.

In several applications in the literature Buchert's equations are considered together with a supposed power-law behavior of the backreaction quantities in terms of the volume scale factor $a_{D}(t)$ \cite{Larena:2008be,Buchert:2006ya,Chaplygin:2010,Roukema:2013,Chiesa2014}. On this basis the dynamics may be solved in terms of $a_{D}(t)$.
The possible emergence of an effective bulk viscous pressure through backreaction has been discussed in \cite{GRG16}.
But as already mentioned, it is desirable to establish a direct connection between the averaged variables and the underlying inhomogeneous dynamics.
Assuming the existence of an exact solution at the level with characteristic scale $l_{1}$, the additional terms due to kinematical backreaction and averaged curvature on the level with characteristic scale $l_{2}\gg l_{1}$ are then, in principle, directly calculable.
We exemplify this strategy here on the basis of the LTB solution for dust in the hyperbolic case, using a small-curvature approximation with a generally inhomogeneous bang time.
We do not specify from the outset the size of the averaging domain. One expects that it should be of the order of the homogeneity scale.
As a result we obtain the time dependence of the effective scale factor and we quantify the difference to the pure dust case of an EdS universe.
The volume scale factor $a_{D}(t)$ is directly obtained from the expression for the averaging volume which we assume to be a sphere of radius $r_{D}$.
Also the average curvature and the kinematical backreaction are directly and independently calculated from the mentioned LTB solution.
Buchert's equation are then not, as for power-law ansatzes for the average curvature and the kinematical backreaction, equations to determine $a_{D}(t)$ but they become identities or consistency relations.
The curvature function of the LTB metric, taken at the radius of the averaging region, directly determines the average curvature quantity in Buchert's equations.
The LTB curvature function also determines the deviation of the volume scale factor from the scale factor of the pure dust case.
We demonstrate that the kinematic backreaction is zero in linear order in the curvature unless the bang time is inhomogeneous.
Also in second order in the curvature the backreaction vanishes for a homogeneous bang time.
In our simple linearized configuration
the effective Hubble rate in terms of the volume scale factor is found to be of the structure of the simplest
phenomenological solution of Buchert's equations in which the parameters are given by the quantities  of the underlying LTB solution at the surface of the averaging region.

To relate this formalism to observations, a further ingredient is required. The averaging procedure over spacelike hypersurfaces leaves open the question of how light propagates.
There does not exist a space-time metric to which the usual condition $ds^{2} =0$ for light propagation could be applied.
A provisional way to handle this problem has been to assume the existence of a template metric \cite{ParanjapeSingh:2006,Larena:2008be} in which the volume scale factor is supposed to play the same r\^{o}le as the scale factor does in the metric of FLRW models, although this template metric is not required to be a solution of the field equations.
Moreover, one assumes that the averaged curvature can be described by a curvature term in the template metric. On this basis several standard techniques for observational tests can be adapted to the averaged dynamics.

 While our LTB based model may be too simple for an adequate description of the real Universe, we believe that it is nevertheless useful as an exactly solvable toy model.

The structure of the paper is as follows. In section \ref{buchert} we summarize the basic relations of Buchert's approach which are given an effective fluid description in section \ref{fluid}. Section \ref{LeTol} recalls basic properties of the spherically-symmetric LTB solution.
Spatial volume averages over the relevant scalars of the LTB solution are considered in section \ref{averaging}. The small-curvature solution of the LTB dynamics is found in section \ref{solutions}. On this basis we calculate and discuss the averaged quantities in section \ref{averaged dynamics} which includes our main results.
Section \ref{effectivemetric} uses the concept of a template metric to make contact with observations of supernovae of type Ia and of the volume expansion rate.
In section \ref{models} we discuss simple models of the LTB curvature function and the bang-time function.
A summary of the paper is given in section \ref{summary}.
%%%%%%%%%%%%%%%%%%%%%%%%%%%%%%%%%%%%%%%%%%%%%%%%%%%%%%%%%%%%%%
\section{The Buchert equations}
\label{buchert}

We start by recalling Buchert's equations for irrotational dust \cite{Buchert:1999er}. These are
\begin{equation}\label{Ftype}
\left(\frac{\dot{a}_{D}}{a_{D}}\right)^{2}
- \frac{8\pi G}{3}\left\langle\rho_{m}\right\rangle_{D} = - \frac{\mathcal{R}_{D} + \mathcal{Q}_{D}}{6}
\end{equation}
and
\begin{equation}\label{ddotb}
\frac{\ddot{a}_{D}}{a_{D}} + \frac{4\pi G}{3}\left\langle\rho_{m}\right\rangle_{D} = \frac{\mathcal{Q}_{D}}{3},
\end{equation}
together with the matter conservation
\begin{equation}\label{mbalD}
\left\langle\rho_{m}\right\rangle_{D}^{\displaystyle\cdot} + 3 \frac{\dot{a}_{D}}{a_{D}}\left\langle\rho_{m}\right\rangle_{D} = 0.
\end{equation}
In these equations $\rho_{m}$ denotes the matter density (irrotational dust), $\Theta$ is the expansion scalar,  $\mathcal{Q}_{D}$ is the kinematical backreaction
\begin{equation}\label{QD}
\mathcal{Q}_{D} = \frac{2}{3}\left\langle\left(\Theta - \left\langle\Theta\right\rangle_{D}\right)^{2}\right\rangle_{D}
- 2 \left\langle\sigma^{2}\right\rangle_{D}\
\end{equation}
and $\mathcal{R}_{D}$ is the averaged three curvature of the time-orthogonal hypersurfaces $t=$ const,
\begin{equation}\label{RD}
\mathcal{R}_{D} =\left\langle ^{3}R\right\rangle_{D}.
\end{equation}
The averages in these equations are volume averages of scalar quantities $S(t,r)$ over a rest mass preserving domain $D$ of volume $V_{D}$  on hypersurfaces $t= $ constant:
\begin{equation}\label{avSgen}
\left\langle S\right\rangle_{D} = \frac{1}{V_{D}}\int_{D}S(t,r)\sqrt{|g_{ij}|}d^{3}r , \qquad V_{D} = \int_{D}\sqrt{|g_{ij}|}d^{3}r ,
\end{equation}
where $|g_{ij}|$ is the determinant of the spatial three-metric on time-orthogonal hypersurfaces.
So far, the size and the structure of the domain $D$ are not specified, it is only assumed that
the evolution of the dust configuration is nonsingular, something which is not
necessarily guaranteed.
The volume scale factor $a_{D}(t)$ is defined by
\begin{equation}\label{aD}
a_{D}(t) = \left[\frac{V_{D}(t)}{V_{D0}}\right]^{1/3} ,
\end{equation}
where $V_{D0}= V_{D}(t_{0})$ is a reference volume of the domain $D$ at a time $t_{0}$.
Despite of the formal similarity to the basic equations of an FLRW universe it should be emphasized that the domain-dependent volume scale factor $a_{D}(t)$ is not the scale factor of a RW metric.
The quantities $\mathcal{Q}_{D}$ and $\mathcal{R}_{D}$ are related by the consistency condition
\begin{equation}\label{intcond}
\frac{1}{a_{D}^{6}}\left(\mathcal{Q}_{D}a_{D}^{6}\right)^{\displaystyle\cdot}
+ \frac{1}{a_{D}^{2}}\left(\mathcal{R}_{D}a_{D}^{2}\right)^{\displaystyle\cdot} = 0.
\end{equation}
The set of equations (\ref{Ftype}) - (\ref{intcond})  is rather general, it is derived by using nothing but the 3+1 decomposition of Einstein's equations together with the matter model of irrotational dust \cite{Buchert:1999er}.
In the simplest case relation (\ref{intcond}) is satisfied by
\begin{equation}\label{QRsimple}
\mathcal{Q}_{D} \propto a_{D}^{-6},\qquad \mathcal{R}_{D} \propto a_{D}^{-2}.
\end{equation}
Below we shall return to this solution in a LTB context.
\noindent
Use of the quantities
\begin{equation}\label{defOmega}
\Omega_{m}^{D} = \frac{8\pi G}{3\mathcal{H_{D}}^{2}}\left\langle\rho_{m}\right\rangle_{D}, \quad
\Omega_{Q}^{D} = - \frac{\mathcal{Q}_{D}}{6\mathcal{H_{D}}^{2}}, \quad
\Omega_{\mathcal{R}}^{D} = - \frac{\mathcal{R}_{D}}{6\mathcal{H_{D}}^{2}},\qquad
\mathcal{H_{D}} = \frac{\dot{a}_{D}}{a_{D}},
\end{equation}
where $\mathcal{H_{D}}$ is the effective Hubble rate,
transforms the Friedmann-type equation (\ref{Ftype}) into
\begin{equation}\label{sum}
\Omega_{m}^{D} + \Omega_{Q}^{D} + \Omega_{R}^{D} = 1.
\end{equation}

\section{Effective fluid description}
\label{fluid}

According to \cite{buchert07}, kinematic backreaction and averaged curvature may be interpreted in terms of an effective backreaction fluid by (the subindex b denotes backreaction)
\begin{equation}\label{rhob}
\rho_{bD} = - \frac{1}{16\pi G}\left(\mathcal{Q}_{D} + \mathcal{R}_{D}\right)\ , \qquad p_{bD} = - \frac{1}{16\pi G}\left(\mathcal{Q}_{D} - \frac{\mathcal{R}_{D}}{3}\right),
\end{equation}
where $\rho_{bD}$ is an effective energy density and $p_{bD}$ is an effective pressure.
With the definitions (\ref{rhob}) the equations (\ref{Ftype}) and (\ref{ddotb}) can be rewritten in the Friedmann-type form
\begin{equation}\label{frb}
\left(\frac{\dot{a}_{D}}{a_{D}}\right)^{2}
- \frac{8\pi G}{3}\left(\left\langle\rho_{m}\right\rangle_{D} + \rho_{bD}\right) = 0,
\end{equation}
\begin{equation}\label{ddotb2}
\frac{\ddot{a}_{D}}{a_{D}} + \frac{4\pi G}{3}\left(\left\langle\rho_{m}\right\rangle_{D} + \rho_{bD} + 3 p_{bD}\right) = 0,
\end{equation}
which implies the conservation law
\begin{equation}\label{consb}
\dot{\rho}_{bD} + 3 \frac{\dot{a}_{D}}{a_{D}}\left(\rho_{bD} + p_{bD}\right) = 0 \
\end{equation}
for the backreaction fluid.
One may define a total energy density $\rho_{D}$,
\begin{equation}\label{rhoT}
\rho_{D} = \left\langle\rho_{m}\right\rangle_{D} + \rho_{bD},
\end{equation}
together with a total pressure $p_{D} \equiv p_{bD}$ which obey the conservation equation
\begin{equation}\label{consrhoT}
\dot{\rho}_{D} + 3 \frac{\dot{a}_{D}}{a_{D}}\left(\rho_{D} + p_{D}\right) = 0.
\end{equation}
The effective equation-of-state (EoS) parameter of the backreaction fluid is
\begin{equation}\label{EoS}
\frac{p_{bD}}{\rho_{bD}} = \frac{\mathcal{Q}_{D} - \frac{1}{3}\mathcal{R}_{D}}{\mathcal{Q}_{D} + \mathcal{R}_{D}}.
\end{equation}
A domain dependent deceleration parameter can be introduced by
\begin{equation}\label{qD}
q_{D} \equiv - \frac{\ddot{a}_{D}a_{D}}{\dot{a}_{D}^{2}},
\end{equation}
which in terms of $\Omega_{Q}^{D}$  and $\Omega_{R}^{D}$ can be written as
\begin{equation}\label{qDO}
q_{D} = \frac{1}{2} + \frac{3}{2}\Omega_{Q}\left[1 - \frac{1}{3}\frac{\Omega_{R}}{\Omega_{Q}}\right].
\end{equation}
This setup is completely general, in particular, no symmetry assumption has been made so far.
In the following section we consider the spherically symmetric LTB dynamics which subsequently will be used
to exemplify the averaging procedure introduced in the previous section.

 %%%%%%%%%%%%%%%%%%%%%%%%%%%%%%%%%%%%%%%%%%%%%%%%%%%%%%%%%%%%%%%%%%%%%%%%%%%%%%%%%%%%%%%

%%%%%%%%%%%%%%%%%%%%%%%%%%%%%%%%%%%%%%%%%%%%%%%%%%%%%%%%%%%%%%

\section{Lema\^{\i}tre-Tolman-Bondi (LTB) dynamics}
\label{LeTol}

As the simplest inhomogeneous dynamics we consider the spherically symmetric LTB solution for irrotational dust (see, e.g., \cite{PleKra}),
\begin{equation}\label{Kmetric}
ds^{2} = dt^{2} - \frac{R^{\prime\, 2}}{1 + 2E(r)}dr^{2} - R^{2}(t,r)\left[d\vartheta^{2} + \sin^{2}\vartheta d\varphi^{2}\right]\,,
\end{equation}
where the function $R = R(t,r)$ obeys
\begin{equation}\label{Kdr2}
\dot{R}^{2} = 2E(r) + \frac{2M(r)}{R}, \qquad \frac{\ddot{R}}{R} = - \frac{M}{R^{3}}.
\end{equation}
For the matter density $\rho_{m}$ one has
\begin{equation}\label{Mpr2}
8\pi G\,\rho_{m} = \frac{2M^{\prime}}{R^{2}R^{\prime}},
\end{equation}
where the prime denotes a derivative with respect to $r$.
The generally valid relation
\begin{equation}\label{}
\frac{1}{3}\Theta^{2} -\sigma^{2} = 8\pi G\rho_{m} - \frac{1}{2}\,^{3}R
\end{equation}
is satisfied in our case with  the expansion scalar $\Theta$,
\begin{equation}\label{Theta}
\Theta = 2\frac{\dot{R}}{R} + \frac{\dot{R}^{\prime}}{R^{\prime}}\,,
\end{equation}
the square of the shear $\sigma$, defined by
\begin{equation}\label{shear}
\sigma^{2} = \frac{1}{3}\left(\frac{\dot{R}^{\prime}}{R^{\prime}} - \frac{\dot{R}}{R}\right)^{2}
\end{equation}
and the three-curvature scalar of the LTB metric
\begin{equation}\label{}
^{3}R = - 4 \frac{\left(ER\right)^{\prime}}{R^{2}R^{\prime}}.
\end{equation}
The matter density obeys the conservation law
\begin{equation}\label{dotrho}
\dot{\rho}_{m} + \Theta\rho_{m} = 0\,.
\end{equation}
There exists a large body of literature on cosmological models relying on the LTB dynamics, see, e.g.,
\cite{tomita99,cel99,iguchi,mustapha,cel12,marra,celerierbolejkokra,sundell,sundell2}.
Our interest here is not primarily whether or not these inhomogeneous models can provide viable alternatives
to the standard $\Lambda$CDM model. Our focus is on the homogeneous average of the inhomogeneous solutions which are supposed to result in a modified dynamics compared with the  dynamics of the standard model, the latter starting with the homogeneity assumption from the outset.

\section{Averaging the LTB scalars}
\label{averaging}

To combine the LTB dynamics with the Buchert equations one needs to consider
the scalars of expansion (\ref{Theta}) and shear (\ref{shear}) in the general expressions  (\ref{QD}) and (\ref{RD}).
The LTB volume element is
\begin{equation}\label{}
d^{3}r = \frac{R^{\prime}R^{2}\sin\vartheta}{\sqrt{1 + 2E}}\,drd\vartheta d\varphi\,.
\end{equation}
Assuming the averaging volume to be a sphere of radius $r_{D}$, this volume becomes
\begin{equation}\label{VD}
V_{D} = 4\pi \int_{0}^{r_{D}}\frac{R^{\prime}R^{2}}{\sqrt{1 + 2E}}\,dr
= \frac{4\pi}{3} \int_{0}^{r_{D}}\frac{\partial}{\partial r}\left(R^{3}\right)\frac{1}{\sqrt{1 + 2E}}.
\end{equation}
The extent of the averaging volume, i.e. the size of the radius $r_{D}$ remains still unspecified here.
Average values of any scalar $S$ are then calculated according to
\begin{equation}\label{avS}
\left\langle S\right\rangle(t) = \frac{4\pi}{V_{D}}\int_{0}^{r_{D}}S(t,r)\,\frac{R^{\prime}(t,r)R^{2}(t,r)}{\sqrt{1 + 2E(r)}}\,dr.
\end{equation}
The combination $\frac{2}{3}\Theta^{2} - 2\sigma^{2}$ which appears in the expression (\ref{QD}) is conveniently written as
\begin{equation}\label{}
\frac{2}{3}\Theta^{2} - 2\sigma^{2} = 4 \frac{\dot{R}}{R}\frac{\dot{R}^{\prime}}{R^{\prime}}
+ 2 \frac{\dot{R}^{2}}{R^{2}} = \frac{2}{R^{2}R^{\prime}}\frac{\partial}{\partial r}\left(\dot{R}^{2}R\right).
\end{equation}
The expansion scalar may also be written in terms of a derivative,
\begin{equation}\label{}
\Theta = \frac{2\dot{R}}{R} + \frac{\dot{R}^{\prime}}{R^{\prime}}
= \frac{1}{R^{2}R^{\prime}}\frac{\partial}{\partial r}\left(R^{2}\dot{R}\right).
\end{equation}
Then we may write
\begin{equation}\label{avTsigma}
\left\langle \frac{2}{3}\Theta^{2} - 2\sigma^{2}\right\rangle
=\frac{8\pi}{V_{D}}\int_{0}^{r_{D}}\frac{\partial}{\partial r}\left(\dot{R}^{2}R\right)\,\frac{1}{\sqrt{1 + 2E}}\,dr
\end{equation}
and
\begin{equation}\label{avT}
\left\langle \Theta\right\rangle
= \frac{4\pi}{V_{D}}\int_{0}^{r_{D}}\frac{\partial}{\partial r}\left(\dot{R}R^{2}\right)\,\frac{1}{\sqrt{1 + 2E}}\,dr\,
\end{equation}
to obtain the combination
\begin{equation}\label{Qcomb}
\mathcal{Q}_{D} = \left\langle \frac{2}{3}\Theta^{2} - 2\sigma^{2}\right\rangle - \frac{2}{3}\left\langle \Theta\right\rangle^{2}.
\end{equation}
For the average of the curvature scalar we have
\begin{equation}\label{avR}
\left\langle ^{3}R\right\rangle = -\frac{16\pi}{V_{D}}\int_{0}^{r_{D}}\frac{\partial}{\partial r}\left(ER\right)
\frac{1}{\sqrt{1 + 2E}}\,dr .
\end{equation}
The expressions (\ref{VD}), (\ref{avTsigma}), (\ref{avT}), (\ref{Qcomb}) and (\ref{avR}) are convenient starting points for the explicit calculations of the relevant averages.
After partial integration the volume (\ref{VD}) becomes
\begin{equation}\label{VDgen}
V_{D}
= \frac{4\pi}{3} \frac{R^{3}(t,r_D)}{\sqrt{1 + 2E(r_{D})}}\left[1
+ \frac{\sqrt{1+2E(r_{D})}}{R^{3}(t,r_{D})}\int_{0}^{r_{D}}R^{3}(t,r)\frac{E^{\prime}(r)}{\left(1 + 2E(r)\right)^{3/2}}\,dr \right].
\end{equation}
The functions $R$ and $E$ outside the integral have to be taken at $r=r_{D}$.
The volume scale factor then is
\begin{equation}\label{aDint}
a_{D}(t) = \frac{R(r_{D},t)}{R_{0}(r_{D})}
\left[\frac{1 + \frac{\sqrt{1+2E}}{R^{3}}\int_{0}^{r_{D}}R^{3}\frac{E^{\prime}}{\left(1 + 2E\right)^{3/2}}\,dr}
{1 + \frac{\sqrt{1+ 2E}}{R_{0}^{3}}\int_{0}^{r_{D}}R_{0}^{3}\frac{E^{\prime}}{\left(1 + 2E\right)^{3/2}}\,dr}
\right]^{1/3}\,.
\end{equation}
In the limit $E=0$ the volume scale factor is given by the LTB function $R$, taken at $r = r_{D}$.
For the volume expansion we find
\begin{equation}\label{}
\frac{\dot{a}_{D}}{a_{D}}
= \frac{\dot{R}(r_{_{D}})}{R(r_{D})}\,\frac{1 + \frac{\sqrt{1+2E}}{\dot{R}R^{2}}\int_{0}^{r_{D}}\dot{R}R^{2}\frac{E^{\prime}}{\left(1 + 2E\right)^{3/2}}\,dr}
{1 + \frac{\sqrt{1+2E}}{R^{3}}\int_{0}^{r_{D}}R^{3}\frac{E^{\prime}}{\left(1 + 2E\right)^{3/2}}\,dr}\,.
\end{equation}
For $E=0$ the simplest FLRW limit is recovered for $R = a r$, resulting in $a_{D} = a$ and the volume expansion coincides with the FLRW Hubble rate,
$\frac{\dot{a}_{D}}{a_{D}} = \frac{\dot{a}}{a}$.
Similarly, we find from (\ref{avR})
\begin{equation}\label{RDgen}
\mathcal{R}_{D}= -\frac{8\pi}{V_{D}}\frac{2ER}{\sqrt{1+2E}}
\left[1 + \frac{\sqrt{1+2E}}{2ER}\int_{0}^{r_{D}}2ER\frac{E^{\prime}}{\left(1 + 2E\right)^{3/2}}\,dr\right]
\end{equation}
and from (\ref{Qcomb}) with (\ref{avTsigma}) and (\ref{avT}),
\begin{eqnarray}\label{QDgen}
  \mathcal{Q}_{D} &=& 6\frac{\dot{R}^{2}(r_{D})}{R^{2}(r_{D})}
  \frac{1}{\left[1+ \frac{\sqrt{1+2E}}{R^{3}}\int_{0}^{r_{D}}R^{3}\frac{E^{\prime}}{\left(1 + 2E\right)^{3/2}}dr\right]^{2}}\cdot\nonumber\\
  &&  \cdot \left\{\frac{\sqrt{1+2E}}{\dot{R}^{2}R}\int_{0}^{r_{D}}\dot{R}^{2}R\,\frac{E^{\prime}}{\left(1 + 2E\right)^{3/2}}\,dr + \frac{\sqrt{1+2E}}{R^{3}}\int_{0}^{r_{D}}R^{3}\frac{E^{\prime}}{\left(1 + 2E\right)^{3/2}}\,dr \right.\nonumber\\
  &&\left.- 2\frac{\sqrt{1+2E}}{\dot{R}R^{2}}\int_{0}^{r_{D}}\dot{R}R^{2}\,\frac{E^{\prime}}{\left(1 + 2E\right)^{3/2}}\,dr
  \right.\nonumber\\
&& \left. + \frac{1+2E}{\dot{R}^{2}R^{4}}\left[\int_{0}^{r_{D}}\dot{R}^{2}R\,\frac{E^{\prime}}{\left(1 + 2E\right)^{3/2}}\,dr \right]\left[\int_{0}^{r_{D}}R^{3}\frac{E^{\prime}}{\left(1 + 2E\right)^{3/2}}\,dr\right]\right.\nonumber\\
&& \left.- \frac{1+2E}{\dot{R}^{2}R^{4}}\left[\int_{0}^{r_{D}}\dot{R}R^{2}\,\frac{E^{\prime}}{\left(1 + 2E\right)^{3/2}}\,dr\right]^{2}
\right\}\,.
\label{Qfin}
\end{eqnarray}
Formulas  (\ref{RDgen}) and (\ref{QDgen}) are the most general expressions for the backreaction in the LTB context.
Note that the functions $R$ and $E$ outside the integrals have to be taken at $r=r_{D}$.
One can show explicitly that upon using the LTB equations (\ref{Kdr2})
the set of Buchert's equations is identically satisfied.
With an explicit solution for $R(r,t)$ and a model for $E(r)$ all the averages can, in principle, be calculated.
Obviously, both $\mathcal{R}_{D}$ and $\mathcal{Q}_{D}$ are determined by the parameters of the LTB solution at the boundary of the averaging volume.

%%%%%%%%%%%%%%%%%%%%%%%%%%%%%%%%%%%%%%%%%%%%%%%%%%%%%%%%%%%%%%

\section{Solutions for the LTB dynamics}
\label{solutions}
\subsection{General solutions}
Here we recall the general solutions  of the LTB dynamics. These depend on the sign of the function $E$.
For $E<0$ equation (\ref{Kdr2}) has the solution
\begin{equation}\label{}
R(r,t) =  \frac{M(r)}{-2E(r)}\left(1 -\cos \eta\right)\,,\qquad \eta - \sin\eta = \frac{\left(-2E(r)\right)^{3/2}}{M(r)}
\left(t- t_{B}(r)\right).
\end{equation}
Here appears another free function of $r$, the bang time $t_{B}(r)$.
The solution for $E>0$ is
\begin{equation}\label{solE>}
R(r,t) =  \frac{M(r)}{2E(r)}\left(\cosh \eta -1\right)\,,\qquad  \sinh\eta  - \eta = \frac{\left(2E(r)\right)^{3/2}}{M(r)}
\left(t - t_{B}(r)\right).
\end{equation}
For $E = 0$ equation (\ref{Kdr2}) is solved by
\begin{equation}\label{R}
 R(r,t) = \left[\frac{9}{2}M(r)\right]^{1/3}\left(t - t_{B}(r)\right)^{2/3}.
\end{equation}
In general, the solutions are characterized by the free functions $M(r)$, $E(r)$ and $t_{B}(r)$.

\subsection{Small-curvature solution for  $E>0$}

Let's focus now on the solution (\ref{solE>}).
We start by realizing that in the limit of small $\eta$, i.e.,
$\cosh \eta \approx 1 + \frac{1}{2}\eta^{2}$, $\sinh \eta \approx \eta + \frac{1}{6}\eta^{3}$ the solution (\ref{R}) for
$E=0$ is recovered.
Now let's include the next order in the expansions for the cosh and sinh functions,
\begin{equation}\label{}
\cosh \eta - 1 \approx \frac{1}{2}\eta^{2}\left(1 + \frac{1}{12}\eta^{2}\right)\,,\qquad
\sinh \eta - \eta\approx \frac{1}{6}\eta^{3}\left(1 + \frac{1}{20}\eta^{2}\right).
\end{equation}
Via the steps
\begin{equation}\label{}
\eta^{3}\left(1 + \frac{1}{20}\eta^{2}\right) \approx 6\frac{\left(2E\right)^{3/2}}{M}\left(t - t_{B}\right),\nonumber
\end{equation}
and
\begin{equation}\label{}
\eta^{2}\left(1 + \frac{1}{12}\eta^{2}\right)
\approx 2E \left(1 + \frac{1}{20}\eta^{2}\right)\left(\frac{6}{M}\right)^{2/3}\left(t - t_{B}\right)^{2/3},\nonumber
\end{equation}
it follows that
\begin{equation}\label{}
R(r,t) = \frac{M}{2E}\frac{1}{2}\eta^{2}\left(1 + \frac{1}{12}\eta^{2}\right)
\end{equation}
and, up to second order in $\eta^{2}$, equivalent to linear order in $E$,
\begin{equation}\label{Rnl}
R(r,t) = \left(\frac{9M(r)}{2}\right)^{1/3}\left(t - t_{B}(r)\right)^{2/3}
\left[1 + \frac{9}{20}(2E(r))\left(\frac{2}{9M(r)}\right)^{2/3}\left(t - t_{B}(r)\right)^{2/3}\right]\,.
\end{equation}
This is the solution for the function $R(r,t)$ for $E>0$, linearized about the solution for $E=0$.
It depends on the spatial functions $M(r)$, $E(r)$ and the inhomogeneous bang time $t_{B}(r)$.
A similar solution was found in \cite{biswasMN,biswasN}.
In the limit $E=0$ the solution (\ref{R}) is recovered. One of the three functions $M(r)$, $E(r)$ and $t_{B}(r)$ may be fixed.
Here we make the choice
\begin{equation}\label{M}
M = \frac{2}{9}\frac{r^{3}}{\left(t_{0} - t_{B}(r)\right)^{2}}\,.
\end{equation}
This guarantees $R_{0} \equiv R(r,t_{0})= r$ for $E=0$ and for a constant $t_{B}$ it reduces to the frequently chosen gauge $M\propto r^{3}$.
With (\ref{M}) the solution (\ref{Rnl}) is written as
\begin{equation}\label{Rnlexpl}
R(r,t) = r\frac{\left(t - t_{B}(r)\right)^{2/3}}{\left(t_{0} - t_{B}(r)\right)^{2/3}}
\left[1 + \kappa\left(r\right) \frac{\left(t - t_{B}(r)\right)^{2/3}}{\left(t_{0} - t_{B}(r)\right)^{2/3}}\right],
\end{equation}
with the curvature quantity
\begin{equation}\label{kappa}
\kappa\left(r\right) \equiv \frac{9}{20}(2E(r))\frac{\left(t_{0} - t_{B}(r)\right)^{2}}{r^{^{2}}}.
\end{equation}
It is the solution (\ref{Rnlexpl}) with (\ref{kappa}) on which the further considerations in the present paper will rely.

Linearizing in $\kappa$ in the derivatives and powers of $R(r,t)$ as, e.g,
\begin{equation}\label{R2nl}
R^{2}(r,t) = r^{2}\frac{\left(t - t_{B}(r)\right)^{4/3}}{\left(t_{0} - t_{B}(r)\right)^{4/3}}
\left[1 + 2\kappa\left(r\right) \frac{\left(t - t_{B}(r)\right)^{2/3}}{\left(t_{0} - t_{B}(r)\right)^{2/3}}\right],
\end{equation}
one checks explicitly that equations (\ref{Kdr2}) are satisfied for the solution (\ref{Rnlexpl}).
For the local Hubble rate
we find
\begin{equation}\label{Hnl}
H(r,t) \equiv \frac{\dot{R}(r,t)}{R(r,t)} = \frac{2}{3}\frac{1}{t - t_{B}(r)}
\left[1 + \kappa\left(r\right)\frac{\left(t - t_{B}(r)\right)^{2/3}}{\left(t_{0} - t_{B}(r)\right)^{2/3}}\right].
\end{equation}
The spatial derivative of $R$ becomes
\begin{eqnarray}
\label{comp}
R^{\prime}(r,t) &=& \frac{\left(t - t_{B}(r)\right)^{2/3}}{\left(t_{0} - t_{B}(r)\right)^{2/3}}
\left[1 - \kappa\left(r\right) \frac{\left(t - t_{B}(r)\right)^{2/3}}{\left(t_{0} - t_{B}(r)\right)^{2/3}}
\left(1 - \frac{rE^{\prime}}{E} + \frac{2}{3}\,r\,t_{B}^{\prime}
\left(\frac{2}{t-t_{B}} +\frac{1}{t_{0}-t_{B}}\right)\right)\right.
\nonumber\\
  &&\left.\qquad\qquad\qquad\qquad\quad
  - \frac{2}{3}\,r\, t_{B}^{\prime}\left(\frac{1}{t-t_{B}} - \frac{1}{t_{0}-t_{B}})\right)\right].
\end{eqnarray}
Notice that $t_{B}^{\prime} > 0$ diminishes the value of $R^{\prime}(r,t)$. Depending on the model this may potentially lead to a shell-crossing singularity $R^{\prime}=0$ for which the energy density (\ref{Mpr2}) diverges.

\section{Averaged dynamics}
\label{averaged dynamics}

Averaging on the basis of the LTB dynamics has attracted considerable interest in the literature \cite{Rasa2004}
 \cite{ParanjapeSingh:2006}\cite{Sussman08}\cite{Mattsson:2010}
\cite{Sussman12}\cite{Sussman13}. A comprehensive analysis has been performed in \cite{Sussman11}.
We start our analysis with the simplest case of vanishing curvature.

\subsection{Zero curvature limit $E=0$}

The averaging volume $V_{D}$ in (\ref{VDgen}) simplifies to
\begin{equation}\label{}
V_{D}
= \frac{4\pi}{3} R^{3}(r_{D})\,\qquad (E=0).
\end{equation}
Further,
\begin{equation}\label{}
\left\langle \frac{2}{3}\Theta^{2} - 2\sigma^{2}\right\rangle
= 6\frac{\dot{R}^{2}(r_{D})}{R^{2}(r_{D})}\,,\qquad (E=0)
\end{equation}
and
\begin{equation}\label{}
\left\langle \Theta\right\rangle = 3\frac{\dot{R}(r_{D})}{R(r_{D})}\quad \Rightarrow\quad
\left\langle \Theta\right\rangle^{2} = 9\frac{\dot{R}^{2}(r_{D})}{R^{2}(r_{D})}\,\qquad (E=0)\,.
\end{equation}
It follows that
\begin{equation}\label{}
\mathcal{Q}_{D} = \frac{2}{3}\left(\left\langle\Theta^{2}\right\rangle - \left\langle \Theta\right\rangle^{2}\right)
- 2\left\langle\sigma^{2}\right\rangle = 0\,\qquad (E=0)\,.
\end{equation}
There is no resulting backreaction for $E=0$ (cf. \cite{ParanjapeSingh:2006}\cite{Mattsson:2010}) and the averaged curvature (\ref{RDgen})
is zero identically.
\ \\
\subsection{Averaged dynamics at first order in the curvature $E$}
Now  we consider the formulas (\ref{VD}), (\ref{avTsigma}), (\ref{avT}), (\ref{Qcomb}) and (\ref{avR}) up to linear order in the curvature function $E$. This requires knowledge of the solution (\ref{Rnlexpl})
for $R(t,r)$ which itself is linear in the curvature.
 On this basis we shall find explicit expressions for the volume scale factor, the effective Hubble rate, the  kinematic backreaction and the average curvature.
\subsubsection{Scale factor}
Linearizing in $E$, the volume expression (\ref{VDgen}) becomes
\begin{equation}\label{}
V_{D} = \frac{4\pi}{3}R^{3}(r_{D})\left[1 - E(r_{D})
+ \frac{1}{R^{3}(r_{D})}\int_{0}^{r_{D}}R^{3}E^{\prime}\,dr\right] + \mathcal{O}(E^{2}).
\end{equation}
With the solution (\ref{Rnlexpl}) at $r=r_{D }$ the volume scale factor then is
\begin{eqnarray}
\label{aDdirect}
a_{D} &=& \frac{\left(t - t_{B}(r_{D})\right)^{2/3}}{\left(t_{0} - t_{B}(r_{D})\right)^{2/3}}\left[1 + \kappa\left(r_{D}\right) \left(\frac{\left(t - t_{B}(r_{D})\right)^{2/3}}{\left(t_{0} - t_{B}(r_{D})\right)^{2/3}}
-1\right)\qquad \qquad\qquad
\right.\nonumber\\
&& \left. \qquad\qquad\qquad + \frac{1}{3R^{3}(r_{D})}\int_{0}^{r_{D}}R^{3}E^{\prime}\,dr
- \frac{1}{3R_{0}^{3}(r_{D})}\int_{0}^{r_{D}}R_{0}^{3}E^{\prime}\,dr\right].
\end{eqnarray}
The LTB curvature modifies the cosmic time dependence of the scale factor compared  with the pure dust case which is recovered for $E=\kappa =0$.
The strength of the modification depends on the value of the curvature parameter $\kappa$ at the border of the domain. The second term in the bracket on the right-hand side of Eq.~(\ref{aDdirect}) induces a faster growth of
the scale factor compared with the  dust universe without backreaction. The additional $t^{2/3}$ dependence coincides exactly with the corresponding dependence found in \cite{LiDominik:2007} on the basis of a perturbation treatment.

\subsubsection{Volume expansion}
The effective Hubble rate $\mathcal{H}_{D}= \frac{1}{3}\frac{\dot{V}_{D}}{V_{D}}$
is determined by
\begin{equation}\label{dV/V}
\frac{\dot{V}_{D}}{V_{D}} = 3 \frac{\dot{R}(r_{D})}{R(r_{D})}\left[1 + \frac{1}{\dot{R}R^{2}}\int_{0}^{r_{D}}\dot{R}R^{2}E^{\prime}\,dr
- \frac{1}{R^{3}(r_{D})}\int_{0}^{r_{D}}R^{3}E^{\prime}\,dr\right] + \mathcal{O}(E^{2}).
\end{equation}
From the linear solution we find for the factor in front of the bracket on the right-hand side of (\ref{dV/V})
(cf. (\ref{Hnl}))
\begin{equation}\label{HnlD}
\frac{\dot{R}(r_{D})}{R(r_{D})} = \frac{2}{3}\frac{1}{t - t_{B}(r_{D})}
\left[1 + \kappa\left(r_{D}\right)\frac{\left(t - t_{B}(r_{D})\right)^{2/3}}{\left(t_{0} - t_{B}(r_{D})\right)^{2/3}}\right].
\end{equation}
Since (cf.(\ref{aDdirect}))
\begin{equation}\label{aD0}
\frac{\left(t - t_{B}(r_{D})\right)^{2/3}}{\left(t_{0} - t_{B}(r_{D})\right)^{2/3}} =
a_{D}\left[1 + \mathcal{O}(E)\right],
\end{equation}
we obtain, up to terms linear in the curvature,
\begin{eqnarray}
\label{calHH0a}
\frac{\mathcal{H}^{2}_{D}}{\mathcal{H}^{2}_{D0}} &=&  a_{D}^{-3}
\left[1 + 5\kappa\left(r_{D}\right)
\left(a_{D} - 1\right)\right. \nonumber\\
  && \left. \qquad\qquad + \frac{2}{\dot{R}(r_{D})R^{2}(r_{D})}\int_{0}^{r_{D}}\dot{R}R^{2}E^{\prime}\,dr
- \frac{1}{R^{3}(r_{D})}\int_{0}^{r_{D}}R^{3}E^{\prime}\,dr\right.\nonumber\\
  && \left. \qquad\qquad - \frac{2}{\dot{R}_{0}(r_{D})R_{0}^{2}(r_{D})}\int_{0}^{r_{D}}\dot{R}_{0}R^{2}_{0}E^{\prime}\,dr
+ \frac{1}{R_{0}^{3}(r_{D})}\int_{0}^{r_{D}}R_{0}^{3}E^{\prime}\,dr\right],
\end{eqnarray}
where $\mathcal{H}_{D0} = \mathcal{H}_{D}(t_{0})$.
The $a_{D}^{-3}$ behavior of the pure dust case is modified accordingly. The curvature term is given as a function of the effective scale factor.

\subsubsection{Kinematical backreaction}
In linear order in $E$, with
\begin{equation}\label{}
\frac{\dot{R}^{2}(r_{D})}{R^{2}(r_{D})} = \mathcal{H}^{2}\left[1+\mathcal{O}(E)\right],
\end{equation}
the kinematical backreaction (\ref{QDgen}) reduces to
\begin{equation}\label{QlinH}
 \mathcal{Q}_{D} = 6\mathcal{H}_{D}^{2}\left[\frac{1}{R^{3}}\int_{0}^{r_{D}}R^{3}E^{\prime} \,dr
 + \frac{1}{M}\int_{0}^{r_{D}}M\,E^{\prime}\,dr
- \frac{2}{\dot{R}R^{2}}\int_{0}^{r_{D}}\dot{R}R^{2}E^{\prime}\,dr\right].
\end{equation}
The kinematical backreaction parameter $\Omega_{Q}^{D}$ becomes
\begin{equation}\label{defOmegaQ}
\Omega_{Q}^{D} = - \frac{\mathcal{Q}_{D}}{6\mathcal{H_{D}}^{2}} = \frac{2}{\dot{R}R^{2}}\int_{0}^{r_{D}}\dot{R}R^{2}E^{\prime}\,dr
- \frac{1}{M}\int_{0}^{r_{D}}M\,E^{\prime}\,dr
- \frac{1}{R^{3}}\int_{0}^{r_{D}}R^{3}E^{\prime} \,dr .
\end{equation}
Explicitly,
\begin{eqnarray}\label{OQexpl}
\Omega_{Q}^{D} &=& \frac{\left(t_{0}- t_{B}(r_{D})\right)^{2}}{r^{3}_{D}}
\left[\frac{2}{t - t_{B}(r_{D})}\int_{0}^{r_{D}}r^{3}\frac{t-t_{B}(r)}{\left(t_{0}- t_{B}(r)\right)^{2}}E^{\prime}dr \qquad\qquad
\right.\nonumber\\
&&\left. \qquad \quad - \frac{1}{\left(t- t_{B}(r_{D})\right)^{2}}\int_{0}^{r_{D}}r^{3}\frac{\left(t-t_{B}(r)\right)^{2}}{\left(t_{0}- t_{B}(r)\right)^{2}}E^{\prime}dr
 - \int_{0}^{r_{D}}\frac{r^{3}}{\left(t_{0}- t_{B}(r)\right)^{2}}E^{\prime}dr\right].
\end{eqnarray}
From the structure of (\ref{OQexpl}) it is obvious that there is no resulting kinematic backreaction for a homogeneous bang time $t_{B}(r) = $ constant. For a constant $t_{B}(r)$ the integrals in (\ref{OQexpl}) just cancel.

\subsubsection{Average curvature}
The linear-order averaged curvature is
\begin{equation}\label{Rava}
\mathcal{R}_{D}
= - 6 \frac{2E(r_{D})}{R^{2}(r_{D})}
= -6 \frac{2E(r_{D})}{r_{D}^{2}a_{D}^{2}}.
\end{equation}
This corresponds to an effective curvature constant $\mathcal{K}_{D}$,
\begin{equation}\label{RcD}
|\mathcal{K}_{D}| = \frac{2E(r_{D})}{r_{D}^{2}} = \mathcal{R}^{-2}_{cD}\quad \Rightarrow\quad
\mathcal{R}_{cD} = \frac{r_{D}}{\sqrt{2E}},
\end{equation}
where $\mathcal{R}_{cD}$ is the effective curvature radius.
This has the structure of a usual curvature term in Friedmann's equation.
In particular, $\mathcal{K}_{D}$ is constant. But the curvature term here is the result of an averaging procedure and it is determined by the parameters of the underlying LTB solution at the border of the averaging volume.
The corresponding curvature parameter reduces to
\begin{equation}\label{ORexpla}
\Omega_{R}^{D}= \frac{9}{4}(2E(r_{D}))\frac{\left(t_{0}- t_{B}(r_{D})\right)^{2}}{r^{2}_{D}}a_{D}
= 5\kappa\left(r_{D}\right)a_{D}.
\end{equation}
This implies $\Omega_{R0}^{D} = 5\kappa\left(r_{D}\right)$ for  $\Omega_{R0}^{D} =  \Omega_{R}^{D}(t_{0})$, its value at $t_{0}$.
For the  curvature radius we have
\begin{equation}\label{RcDH}
\mathcal{R}_{cD} = \frac{c}{\mathcal{H}_{D0}\sqrt{\Omega_{R0}^{D}}}\,,
\end{equation}
i.e., it is of the order of the Hubble radius of the domain which we shall assume now to be the
entire observable Universe.

\subsubsection{Matter fraction}
The behavior $\left\langle \rho_{m}\right\rangle_{D} \propto a_{D}^{-3}$ which is a consequence of the conservation equation (\ref{mbalD}), is consistent with the average of (\ref{Mpr2}). Namely,
\begin{equation}\label{}
8\pi G\,\left\langle\rho_{m} \right\rangle_{D} = \frac{6M(r_{D})}{R^{3}(r_{D})}\,\left[1 + \frac{1}{M(r_{D})}\int^{r_{D}}_{0}dr\,M(r)E^{\prime}
- \frac{1}{R^{3}(r_{D})}\int^{r_{D}}_{0} dr R^{3}E^{\prime}
\right].
\end{equation}
The density ratio $\frac{\left\langle\rho_{_{m}} \right\rangle_{D}}{\left\langle\rho_{m} \right\rangle_{D0}}$ then correctly becomes
\begin{equation}\label{}
\frac{\left\langle\rho_{_{m}} \right\rangle_{D}}{\left\langle\rho_{m} \right\rangle_{D0}}
= \frac{R_{0}^{3}}{R^{3}}\left[1
- \frac{1}{R^{3}(r_{D})}\int^{r_{D}}_{0} dr R^{3}E^{\prime} + \frac{1}{R_{0}^{3}(r_{D})}\int^{r_{D}}_{0} dr R_{0}^{3}E^{\prime}
\right] = a_{D}^{-3}.
\end{equation}
The expression for the matter density parameter is
\begin{equation}\label{}
\Omega_{m} = \frac{2M}{R^{3}}\frac{R^{2}}{\dot{R}^{2}}
\left[1 + \frac{1}{M}\int_{0}^{r_{D}}ME^{\prime} \,dr -  \frac{2}{\dot{R}R^{2}}\int_{0}^{r_{D}}\dot{R}R^{2}E^{\prime}\,dr
+ \frac{1}{R^{3}(r_{D})}\int^{r_{D}}_{0} dr R^{3}E^{\prime}\right].
\end{equation}
Up to linear order
\begin{equation}\label{}
R^{3}\frac{\dot{R}^{2}}{R^{2}}= 2M\left[1 + 5\kappa\left(r_{D}\right)a_{D}
\right] + \mathcal{O}(E^{2})
\end{equation}
is valid and with (\ref{defOmegaQ}) and (\ref{ORexpla}) one verifies that (\ref{sum}) is consistently recovered at this order.

\subsubsection{Consistency}
Obviously, $\left(\mathcal{R}_{D}a_{D}^{2}\right)^{\displaystyle\cdot} = 0$. Then the consistency relation (\ref{intcond}) dictates
that either $\mathcal{Q}_{D} = 0$ or  $\mathcal{Q}_{D} \propto a_{D}^{-6}$.
By direct calculation one verifies that indeed
\begin{equation}\label{QRexpla}
Q_{D} = Q_{D0}a_{D}^{-6},\quad \Omega_{Q}^{D} = \Omega_{Q0}^{D}a_{D}^{-3}
\end{equation}
with
\begin{eqnarray}\label{OQexpl0}
\Omega_{Q0}^{D} &=& \frac{1}{r^{3}_{D}}
\left[2\left(t_{0}- t_{B}(r_{D})\right)\int_{0}^{r_{D}}\frac{r^{3}}{t_{0}- t_{B}(r)}E^{\prime}dr
\right.\nonumber\\
&&\left. \qquad  - \int_{0}^{r_{D}}r^{3}E^{\prime}dr \
 - \ \left(t_{0}- t_{B}(r_{D})\right)^{2}\int_{0}^{r_{D}}\frac{r^{3}}{\left(t_{0}- t_{B}(r)\right)^{2}}E^{\prime}dr\right].
\end{eqnarray}
This combination vanishes for a homogeneous bang time. To have a nonvanishing kinematic backreaction at linear order, an inhomogeneous bang time is necessarily required. For a homogeneous bang time the effective EoS parameter (\ref{EoS}) is always $\frac{p_{bD}}{\rho_{bD}} =  -\frac{1}{3}$.

\subsubsection{Hubble rate and deceleration parameter}
With the parameters (\ref{ORexpla}) and (\ref{QRexpla}) the Hubble rate (\ref{calHH0a}) is written as
\begin{equation}\label{HDa}
\frac{\mathcal{H}^{2}_{D}}{\mathcal{H}^{2}_{D0}} =  a_{D}^{-3}
\left[\Omega_{m0}^{D} + \Omega_{R0}^{D}a_{D} + \Omega_{Q0}^{D}a_{D}^{-3}\right],
\end{equation}
while the deceleration parameter (\ref{qDO}) becomes
\begin{equation}\label{qDa}
q_{D} = \frac{1}{2} + \frac{3}{2}\left[\Omega_{Q0}^{D}a_{D}^{-3} - \frac{1}{3}\Omega_{R0}^{D}a_{D}\right].
\end{equation}
The result (\ref{HDa}) for the Hubble rate in terms of the volume scale factor with explicitly known coefficients
$\Omega_{m0}^{D}$, $\Omega_{R0}^{D}$ and $\Omega_{Q0}^{D}$ is our main achievement so far.
The combination $\Omega_{R0}^{D}a_{D} + \Omega_{Q0}^{D}a_{D}^{-3}$ represents the influence of the backreaction fluid on the dynamics. From the point of view of backreaction cosmology it is supposed to be the equivalent of the dark-sector components in the cosmological standard model. Given the dependence of this contribution on $a_{D}$ it is not obvious, however, that this expectation can be  realized within our simple LTB model.
While the structure of (\ref{HDa}) corresponds to the simplest possible phenomenological solution (\ref{QRsimple}) for which the kinematical backreaction and averaged curvature  terms in the consistency relation (\ref{intcond}) separately vanish, we have derived this structure here from an underlying exact inhomogeneous dynamics
which provided us with explicit expressions for $\Omega_{R0}^{D}$ and $\Omega_{Q0}^{D}$.
Even if it may not lead to a realistic description of our Universe, we believe it to be useful as an exactly solvable toy model and a first step to more realistic configurations.

For a homogeneous bang time the last term in (\ref{HDa}) vanishes since $\Omega_{Q0} = 0$ and the only additional contribution
from the averaging procedure  is due to a constant curvature in which the curvature constant is determined by the LTB solution at the border of the averaging area. The backreaction fluid becomes a pure curvature component in this case.
The emergence of a spatial curvature term as the result of the averaging  procedure is in accord with a corresponding result from Macroscopic Gravity \cite{coley,coley2,coley3}.
For an inhomogeneous bang time the kinematic backreaction is generally different from zero. But since its contribution relative to the matter part decays with $a_{D}^{-3}$ in (\ref{HDa}), it will have a decreasing impact on the dynamics as $a_{D}$ increases. Moreover, since one expects the matter part $\Omega_{m}^{D}$ to dominate at $a_{D}\ll 1$, we have a strong constraint on the current backreaction parameter $\Omega_{Q0}^{D}$.
The age of such kind of universe has to be calculated from
\begin{equation}\label{}
  t_{0} - t_{B}(r_{D}) = \int_{0}^{1} \frac{da_{D}}{a_{D}}\frac{1}{\mathcal{H}_{D0}\sqrt{\Omega_{m0}^{D}a_{D}^{-3} + \Omega_{R0}^{D}a_{D}^{-2}+ \Omega_{Q0}^{D}a_{D}^{-6}}}.
\end{equation}
The backreaction part increases with decreasing $a_{D}$, i.e., towards the past, which is obviously against its expected r\^{o}le for the cosmological dynamics.

We have assumed here an idealized description of the inhomogeneous Universe as one single spherically symmetric
configuration. A more realistic model would have to include a set of different regions with generally different
inhomogeneous distributions. Then, the averages taken here over just one inhomogeneous solution would have to be
performed over an entire set of inhomogeneities.

\ \\
\ \\
\noindent
\subsection{Backreaction at second order}

From the general expressions  (\ref{RDgen}) and (\ref{QDgen}) for the averaged curvature and the kinematical backreaction, respectively, it is obvious that their lowest-order contributions are
at least linear in $E$,  for $E=0$ both $\mathcal{R}_{D}$ and $\mathcal{Q}_{D}$ vanish.
Since we know the solution for $R$ up to linear order as well (cf.(\ref{Rnlexpl})) it is possible to calculate
$\mathcal{R}_{D}$ and $\mathcal{Q}_{D}$ up to second order.
The result for the averaged curvature is
\begin{eqnarray}
\mathcal{R}_{D} &=& - \frac{6}{r_{D}^{2}a_{D}^{2}}\left[
2E\left(1 - \frac{2}{5}\Omega_{R0}^{D} -\frac{2}{3R_{0}^{3}(r_{D})}\int_{0}^{r_{D}}R_{0}^{3}E^{\prime}\,dr\right)  \right.\nonumber\\
 && \left.\qquad\quad +\frac{1}{R(r_{D},t)}\int_{0}^{r_{D}}2E(r)R(r,t)E^{\prime}dr
 - \frac{2E}{3R^{3}(r_{D},t)}\int_{0}^{r_{D}}R^{3}(r,t)E^{\prime}dr \right].
 \label{Rquadr}
\end{eqnarray}
Only the terms of the second line in (\ref{Rquadr}) depend on time. To check the dependence of $\mathcal{R}_{D}$ on $a_{D}$ it is useful to calculate $\left(a_{D}^{2}\mathcal{R}_{D}\right)^{\cdot}$. We obtain
\begin{eqnarray}
\left(a_{D}^{2} \mathcal{R}\right)^{\cdot} &=& -\frac{12\mathcal{H}_{D}(r_{D},t)}{r_{D}^{2}}
\left[E(r_{D})\left(\frac{1}{R^{3}(r_{D},t)}\int_{0}^{r_{D}}R^{3}(r,t)E^{\prime}dr \right.\right.\nonumber\\
&&\left.\left. \qquad\qquad \qquad\qquad -
\frac{1}{\dot{R}(r_{D},t)R^{2}(r_{D},t)}\int_{0}^{r_{D}}\dot{R}(r,t)R^{2}(r,t)E^{\prime}dr\right)\right.\nonumber \\
&& \left. \qquad\qquad - \frac{1}{R(r_{D},t)}\int_{0}^{r_{D}}R(r,t)EE^{\prime}dr
+ \frac{1}{\dot{R}(r_{D},t)}\int_{0}^{r_{D}}\dot{R}(r,t)EE^{\prime}dr\right].
\end{eqnarray}
This is a pure second-order quantity. One can use here the zeroth-order expression
for $R(r,t)$,
\begin{equation}\label{}
R(r,t) = r\frac{\left(t - t_{B}(r)\right)^{2/3}}{\left(t_{0} - t_{B}(r)\right)^{2/3}} + \mathcal{O}(E),
\end{equation}
within the integrals and the same solution at $r=r_{D}$ in the factors that multiply the integrals.
This reveals that for $t_{B}= $constant the first two terms cancel each other and the third and fourth terms cancel each other as well. For a homogeneous bang time the averaged curvature behaves as $a_{D}^{-2}$ even at second order in $E$. Deviations from a constant curvature require an inhomogeneous bang time.
By a straightforward calculation one realizes from (\ref{QDgen}) that $\mathcal{Q}_{D}$ indeed vanishes
also in second order in $E$ for a constant $t_{B}$ which is consistent with the result $\mathcal{R}_{D} \propto a_{D}^{-2}$ for this case.

For our idealized simple LTB configuration a nonvanishing kinematical backreaction can only be realized for a non-simultaneous big bang.
A solution that satisfies relation (\ref{intcond}) beyond the simplest case (\ref{QRsimple}) has to be at least of quadratic order in the LTB curvature parameter $E$ with an inhomogeneous bang time.

\section{Effective metric and luminosity distance}
\label{effectivemetric}

Our formalism so far left open the problem of light propagation  in a backreaction context. The volume scale factor $a_{D}$ is not related to a space-time metric. Here, an additional ingredient is necessary.
To make contact with observations, it is useful to consider an effective metric of the Robertson-Walker type with the  quantity $a_{D}$ as an effective scale factor and (\ref{RcD}) as generalized curvature (cf. \cite{Roukema:2013,GRG16}),
\begin{equation}\label{metriccurv}
ds^{2}_{eff} = c^{2}dt^{2} - a_{D}^{2}\left[dr^{2} + \mathcal{R}_{cD}^{2}\sinh^{2} \frac{r}{\mathcal{R}_{cD}}\left(d\vartheta^{2}
+ \sin^{2}\vartheta d\varphi^{2}\right)\right].
\end{equation}
Under this assumption radial light propagation is described by
\begin{equation}\label{}
ds^{2} = 0 \quad \Rightarrow\quad dr = \pm\frac{c}{a_{D}^{2}\mathcal{H}_{D}}da_{D}
= \mp \frac{c}{\mathcal{H}_{D}}dz_{D},
\end{equation}
where we have introduced an effective redshift parameter $z_{D}$ by $1 + z_{D} = a_{D}^{-1}$.
Then
\begin{equation}\label{rzD}
r(z_{D}) = \frac{c}{\mathcal{H}_{D0}}
\int_{0}^{z_{D}}\frac{dz_{D}}{\left[\Omega_{m0}^{D}(1+z_{D})^{3} + \Omega_{R0}^{D}(1+z_{D})^{2} + \Omega_{Q0}^{D}(1+z_{D})^{6}\right]^{1/2}}.
\end{equation}
The luminosity distance can be calculated via
\begin{equation}\label{dLcurv}
d_{L}^{eff}(z_{D}) = \left(1+z_{D}\right)\mathcal{R}_{cD}(z_{D})\sinh \frac{r(z_{D})}{\mathcal{R}_{cD}(z_{D})}
\end{equation}
with $\mathcal{R}_{cD}$ from (\ref{RcD}).

\begin{figure}[H]
\centering
\includegraphics[scale=0.3]{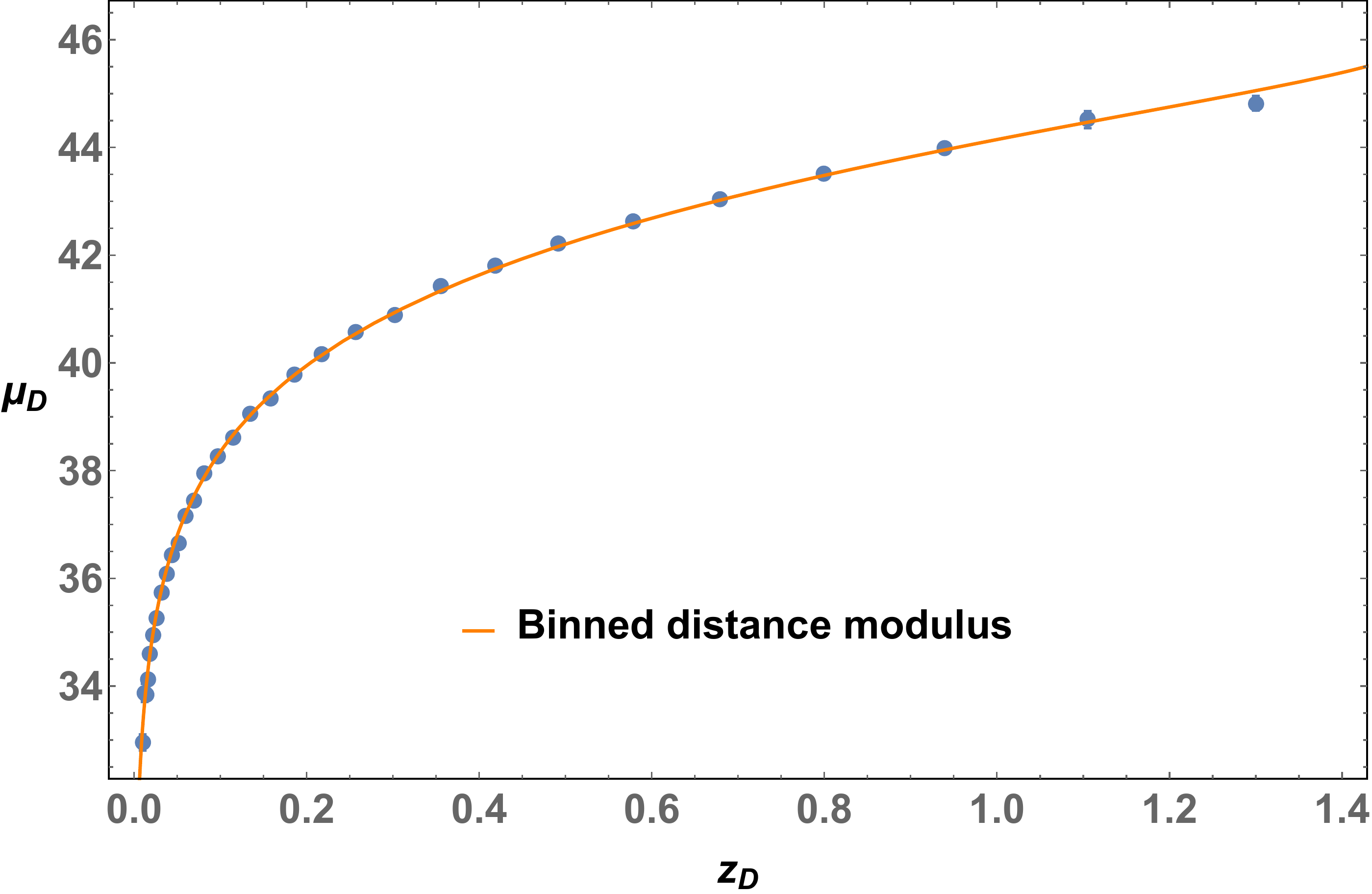}
\caption{Binned distance modulus $\mu_{D}$ in dependence of the redshift parameter $z_{D}$, based on (\ref{dLcurv}).}\label{modulus}
\end{figure}

As already mentioned, the Hubble rate (\ref{HDa}) and, consequently, the expression (\ref{dLcurv}) with (\ref{rzD}) are not expected to result in a competitive model of our real Universe. But even if seen primarily as a toy model, it may be of interest to clarify its status concerning observational data.
We start by a simplified analysis which ignores the detailed structure for the expressions for
 $\Omega_{R0}^{D}$ and $\Omega_{Q0}^{D}$ but allows for a shortcut to observational results.
  To be in rough accord with the standard model, we fix the matter fraction to be $\Omega_{m0}^{D}=0.3$.
 Then we regard
$\Omega_{R0}^{D}$ and $\mathcal{H}_{D0}$ as  free parameters in the expressions (\ref{HDa}) and (\ref{dLcurv})
for $\mathcal{H}_{D}$
and $d_{L}^{eff}(z_{D})$, respectively, and confront  the results with data from supernovae of type Ia, as well as with  differential age data of old galaxies for $\mathcal{H}_{D}(z_{D})$.
Tentatively, we adopt the values from the standard analysis for the distance modulus (cf. \cite{GRG16}),
\begin{equation}
\mu_{D}=5\log d_L^{\mathrm{eff}}(z_{D}) +\mu_{D0}
\label{moduloD}
\end{equation}
with $\mu_{D0}=42.384-5\log h_{D}$, where $h_{D}$ is defined by $\mathcal{H}_{D0} = 100  h_{D} \mathrm{km s^{-1} Mpc^{-1}}$. This choice implies an averaging scale of the order of the size of the observable Universe.
It allows us to perform a statistical analysis using the data from the JLA compilation of type Ia supernovae \cite{betoule}.
The resulting binned distance modulus $\mu_{D}$ in dependence of the redshift parameter $z_{D}$ is shown in Fig.~\ref{modulus}.
%%%%%%%%%%%%%%%%%%%%%%%%%%%%%%%%%%%%%%%%%%%%%%%%%%%%%%%%%%%%%%%%%%%%%%%%%%%%%%%%%%%%%%%%%%%%%%%%%%%%%%%%%%%%%%%%%%%
%%%%%%%%%%%%%%%%%%%%%%%%%%%%%%%%%%%%%%%%%%%%%%%%%%%%%%%%%%%%%%%%%%%%%%%%%%%%%%%%%%%%%%%%%%%%%%%%%%%%%%%%%%%%%%%%%%%%%%
As best-fit values we obtain  $\Omega_{R0}^{D} = 0.74$ and $h_{D}=0.67$.   Using these values in (\ref{RcDH}) we find $\mathcal{R}_{cD} = 5.21 \mathrm{Gpc}$ for the curvature radius of the universe.
The corresponding value for the kinematic backreaction  is $\Omega_{Q0}^{D} = - 0.038$.

\begin{figure}[H]
\centering
\includegraphics[scale=0.4]{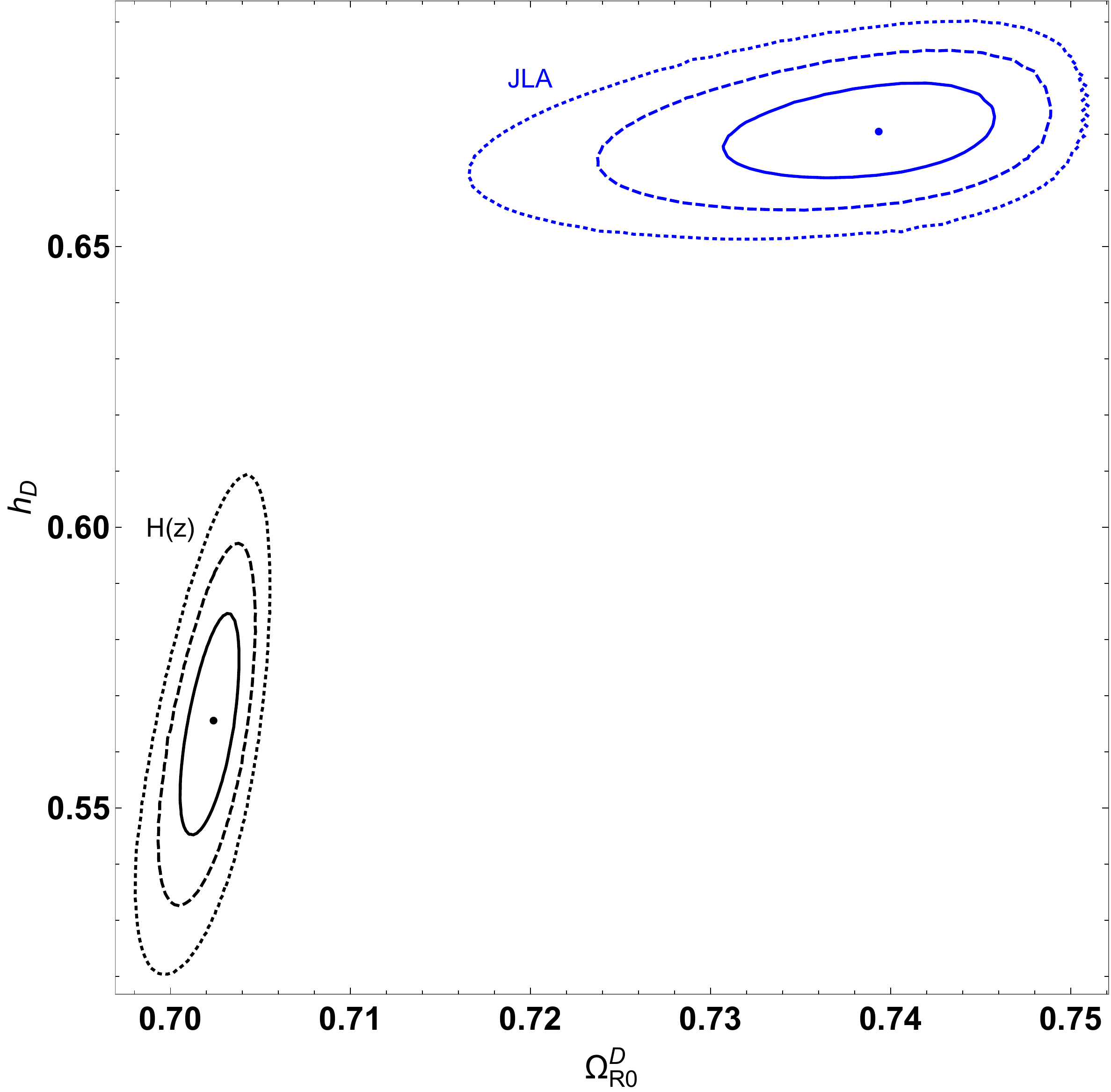}
\caption{The $\Omega_{R0}^{D}$-$h_{D}$ plane with contour plots ($1\sigma$, $2\sigma$ and $3\sigma$) for the SNIa and $\mathcal{H}_{D}(z_{D})$ tests.}\label{plane}
\end{figure}

While this might seem to give some observational support for the model, the situation changes if we apply
a different test which confronts the Hubble rate (\ref{HDa}) with the differential age data of old galaxies that have evolved passively \cite{Ji1, Ji2, Hz}. Here we use the $28$ data points listed in \cite{Farooq}.
This $\mathcal{H}_{D}(z_{D})$ analysis provides us with the rather different values
$\Omega_{R0}^{D} = 0.7023^{+0.0009}_{-0.0012}$  $h=0.5659^{+0.0130}_{}$ and $\Omega_{Q0}^{D} = - 0.002$.
Still more important: although the present values of the fractional kinematic backreaction are very small in both cases,
they are still much too large to allow for a matter dominated phase at redshifts of the order of the redshift of the recombination era.
Moreover, a backreaction which is increasing towards the past at a bigger rate than the matter fraction  is physically doubtful anyway.
The results of the statistical analysis are visualized in Fig.~\ref{plane}.
The confidence contours of both tests are dramatically different which explicitly demonstrates the observational failure of our curvature fluid configuration.
%%%%%%%%%%%%%%%%%%%%%%%%%%%%%%%%%%%%%%%%%%%%%%%%%%%%%%%%%%%%%%%%%%%%%%%%%%%%%%%%%%%%%%%%%%%%%%%%%%%%%%%%%%%%%%%%%%%%%%%%%%
%%%%%%%%%%%%%%%%%%%%%%%%%%%%%%%%%%%%%%%%%%%%%%%%%%%%%%%%%%%%%%%%%%%%%%%%%%%%%%%%%%%%%%%%%%%%%%%%%%%%%%%%%%%%%%%%%%%%%%%%%

Notice that to obtain these results we did not make use of the expressions (\ref{ORexpla}) and  (\ref{OQexpl0}) for
$\Omega_{R0}^{D}$  and $\Omega_{Q0}^{D}$, respectively. We did not require either explicit expressions for the curvature parameter $E(r)$
and for the inhomogeneous bang time $t_{B}(r)$.  In the following section we shall briefly discuss simple models for these quantities.

%%%%%%%%%%%%%%%%%%%%%%%%%%%%%%%%%%%%%%%%%%%%%%%%%%%%%%%%%%%%

\section{Simple models for $E(r)$ and $t_{B}(r)$}
\label{models}

The dynamics described by the Hubble rate (\ref{HDa}) and the luminosity distance (\ref{dLcurv}) which establishes the contact to cosmological observations, depends on the details of the functions $E(r)$ and $t_{B}(r)$ for which specific models have to be implemented.
The function $2E$ has to have the general form (cf.~\cite{PleKra})
\begin{equation}\label{}
2E = - r^{2}\left(k + F(r)\right)\,,\qquad F(0) = 0\,.
\end{equation}
With a choice
\begin{equation}\label{F}
2E = - r^{2}\left(-|k| + F(r)\right)\,,\qquad F = |k|\left(1 - e^{-\left(r/r_{E}\right)^{n}}\right)\,,
\end{equation}
which satisfies $F(0) = 0$ one has
\begin{equation}\label{2E}
2E = r^{2}|k|e^{-\left(r/r_{E}\right)^{n}},\qquad E^{\prime} = \left[1 - \frac{n}{2}\left(\frac{r}{r_{E}}\right)^{n}\right]r|k|e^{-\left(r/r_{E}\right)^{n}}.
\end{equation}
The radius $r_{E}$ characterizes the range of the curvature of the LTB solution.
The curvature function tends to zero in the limit $r\gg r_{E}$.

%\subsubsection{Average curvature radius}
From (\ref{ORexpla}) we have
\begin{equation}\label{ORexplt=0}
\Omega_{R0}^{D}=\frac{9}{4}(2E(r_{D}))\frac{\left(t_{0}- t_{B}(r_{D})\right)^{2}}{r^{2}_{D}}.
\end{equation}
Together with
\begin{equation}\label{}
\mathcal{H}_{D0}^{2} = \frac{4}{9\left(t_{0}- t_{B}(r_{D})\right)^{2}} + \mathcal{O}(E)
\end{equation}
as well as with (\ref{RcD}) and (\ref{RcDH}) we find the correspondence
\begin{equation}\label{rDRD}
2E(r_{D}) = \frac{r_{D}^{2}}{\mathcal{R}_{cD}^{2}}\,, \quad \frac{1}{\mathcal{R}_{cD}^{2}} = |k|e^{-\left(r_{D}/r_{E}\right)^{n}}\,, \quad \Omega_{R0}^{D} = \frac{c^{2}}{\mathcal{H}_{D0}^{2}\mathcal{R}_{cD}^{2}}
.
\end{equation}
The combination $|k|e^{-\left(r_{D}/r_{E}\right)^{n}}$ in the ansatz for $2E(r_{D})$ represents the square of the inverse curvature radius of the averaged dynamics.
The radius $r_{D}$ of the averaging volume has to be smaller than the curvature radius $\mathcal{R}_{cD}^{2}$ to guarantee $2E< 1$, the condition for the applicability of our linear curvature approximation.

If the averaging volume is taken such that $r_{D}\gg r_{E}$ (and assuming a reference value of $k = 1 \mathrm{Gpc}^{-2}$), the curvature radius of the averaged dynamics tends to infinity, i.e., the average curvature is negligible. To have a finite average curvature radius $\mathcal{R}_{cD}$, equivalent to a noticeable influence of the curvature on the average dynamics, the extension $r_{D}$ of the averaging volume has to be of the order of $r_{E}$.
The value
$\mathcal{R}_{cD} = 5.21 \mathrm{Gpc}$ of the JLA test is realized for $r_{D} = 1.79 r_{E}$.

 For a detailed calculation of the quantities $\Omega_{R0}^{D}$ and $\Omega_{Q0}^{D}$  in (\ref{ORexpla}) and (\ref{OQexpl0}), respectively, additionally an explicit model for the inhomogeneous bang time $t_{B}(r)$ is needed.
 In \cite{eddy} it was demonstrated that an ansatz
\begin{equation}\label{tBpr>}
t_{B}(r) = t_{B0}\left(1 - e^{-\left(r/r_{c}\right)^{m}}\right),
\end{equation}
where $r_{c}$ denotes another inhomogeneity scale,
gives rise to a simple void model. The bang time (\ref{tBpr>}) increases with $r$ until it approaches a constant value, i.e.,
\begin{equation}\label{}
t_{B}(0) = 0\,, \qquad t_{B}(r\gg r_{c}) = t_{B0}.
\end{equation}
With these assumptions the integrals in the expression (\ref{OQexpl0}) for the current kinematical backreaction parameter $\Omega_{Q0}^{D}$ may be evaluated explicitly.
Various combinations of the parameters $t_{B0}$, $r_{c}$ and $r_{E}$ were checked, but even though one of them reproduces roughly the same value as the $\mathcal{H}_{D}(z_{D})$ analysis, the physical significance of this term remains doubtful.

\section{Summary}
\label{summary}

We have derived the simplest phenomenological solution of Buchert's equations from an underlying LTB dynamics in the linear curvature approximation. This solution represents an exactly solvable toy model  of how to construct an averaged homogeneous dynamics from an exact inhomogeneous solution.
The averaged variables depend on the parameters of the LTB solution at the boundary of the averaging volume.
For this simple configuration there exists a nonvanishing kinematic backreaction only for an inhomogeneous bang time. This is true both at linear and at quadratic orders in the LTB curvature $E$. For a homogeneous bang time the backreaction fluid is a pure curvature component.
The appearance of an averaging-induced curvature term is also known from Macroscopic Gravity \cite{coley,coley2,coley3}.
At first order in $E$ one has $\mathcal{R}_{D} \propto a_{D}^{-2}$ for the averaged curvature which is similar to the FLRW case. Deviations from this behavior may occur at second order in $E$, but this requires an inhomogeneous bang time.

Both because of its internal dynamics, the kinematic backreaction is growing towards the past,  and because of  its difficulties to account for the present observational data this model does not provide a realistic description of our Universe. It is an exactly solvable toy model which might be a first step towards a better understanding of how a homogeneous and isotropic dynamics could emerge out of an underlying inhomogeneous configuration.

Potential extensions of this study include the investigation of less symmetric models, possibly models based
on the Szekeres metric \cite{szekeres,PleKra}. A more adequate picture will also have to consider the Universe to consist
of more than just one inhomogeneous region. In general, the averaging domain will be made of different and disjoint overdense and underdense regions which will have to be described by different local expansion rates \cite{BuchertRae}.
\\

{\bf Acknowledgement:}  We thank FAPES, CAPES and CNPq (Brazil) for financial support.

%%%%%%%%%%%%%%%%%%%%%%%%%%%%%%%%%%%%%%%%%%%%%%%%%%%%%%%%%%%%%%%%%%%%%%%%%%%%%%%

\end{document}